\documentstyle[12pt]{article}
\newcommand{\nc}{\newcommand}
\nc{\renc}{\renewcommand}

\nc{\half}{{\textstyle{1\over2}}}
\nc{\etal}{\mbox{\it et al. }}
\nc{\ie}{{\it i.e.}}
\nc{\eg}{{\it e.g.}}

\renc{\thefootnote}{\arabic{footnote}}
\nc{\capt}[1]{{\bf Figure.} {\small\sl #1}}

\nc{\eqs}[2]{\mbox{Eqs.~(\ref{#1},\,\ref{#2})}}
\nc{\eq}[1]{\mbox{Eq.~(\ref{#1})}}

\nc{\figs}[2]{\mbox{Figs.~(\ref{#1},\,\ref{#2})}}
\nc{\fig}[1]{\mbox{Fig~.(\ref{#1})}}

\nc{\tag}[1]{\label{#1} \marginpar{{\footnotesize #1}}}
\nc{\mtag}[1]{\label{#1} \mbox{\marginpar{{\footnotesize #1}}}}
\renc{\baselinestretch}{1.5}
\newlength{\overeqskip}
\newlength{\undereqskip}
\nc{\be}[1]{\begin{equation} \mbox{$\label{#1}$}}
\nc{\bea}[1]{\begin{eqnarray} \mbox{$\label{#1}$}}
\nc{\Section}[2]{\section{#2}\label{#1}}
\nc{\Bibitem}[1]{\bibitem{#1}}
\nc{\Label}[1]{\label{#1}}

\nc{\eea}{\vspace{\undereqskip}\end{eqnarray}}
\nc{\ee}{\vspace{\undereqskip}\end{equation}}
\nc{\bdm}{\begin{displaymath}}
\nc{\edm}{\end{displaymath}}
\nc{\dpsty}{\displaystyle}
\nc{\bc}{\begin{center}}
\nc{\ec}{\end{center}}
\nc{\ba}{\begin{array}}
\nc{\ea}{\end{array}}
\nc{\bab}{\begin{abstract}}
\nc{\eab}{\end{abstract}}
\nc{\btab}{\begin{tabular}}
\nc{\etab}{\end{tabular}}
\nc{\bit}{\begin{itemize}}
\nc{\eit}{\end{itemize}}
\nc{\ben}{\begin{enumerate}}
\nc{\een}{\end{enumerate}}
\nc{\bfig}{\begin{figure}}
\nc{\efig}{\end{figure}}
\nc{\arreq}{&\!=\!&}
\nc{\arrmi}{&\!-\!&}
\nc{\arrpl}{&\!+\!&}
\nc{\arrap}{&\!\!\!\approx\!\!\!&}
\nc{\non}{\nonumber\\*}
\nc{\align}{\!\!\!\!\!\!\!\!&&}

\def\lsim{\; \raise0.3ex\hbox{$<$\kern-0.75em
      \raise-1.1ex\hbox{$\sim$}}\; }
\def\gsim{\; \raise0.3ex\hbox{$>$\kern-0.75em
      \raise-1.1ex\hbox{$\sim$}}\; }
\nc{\DOT}{\hspace{-0.08in}{\bf .}\hspace{0.1in}}
\nc{\Laada}{\hbox {$\sqcap$ \kern -1em $\sqcup$}}
\nc\loota{{\scriptstyle\sqcap\kern-0.55em\hbox{$\scriptstyle\sqcup$}}}
\nc\Loota{{\sqcap\kern-0.65em\hbox{$\sqcup$}}}
\nc\laada{\Loota}
\nc{\qed}{\hskip 3em \hbox{\BOX} \vskip 2ex}

\nc{\real}{{\rm I \! R}}
\nc{\Z}{{\sf Z \!\!\! Z}}
\nc{\complex}{{\rm C\!\!\! {\sf I}\,\,}}
\def\bigid{\leavevmode\hbox{\small1\kern-3.8pt\normalsize1}}
\def\id{\leavevmode\hbox{\small1\kern-3.3pt\normalsize1}}
\nc{\slask}{\!\!\!/}
\nc{\bis}{{\prime\prime}}
\nc{\pa}{\partial}
\nc{\na}{\nabla}
\nc{\ra}{\rangle}
\nc{\la}{\langle}
\nc{\goto}{\rightarrow}
\nc{\swap}{\leftrightarrow}

\nc{\EE}[1]{ \mbox{$\cdot10^{#1}$} }
\nc{\abs}[1]{\left|#1\right|}
\nc{\at}[2]{\left.#1\right|_{#2}}
\nc{\norm}[1]{\|#1\|}
\nc{\abscut}[2]{\Abs{#1}_{\scriptscriptstyle#2}}
\nc{\vek}[1]{{\rm\bf #1}}
\nc{\integral}[2]{\int\limits_{#1}^{#2}}
\nc{\inv}[1]{\frac{1}{#1}}
\nc{\dd}[2]{{{\partial #1}\over{\partial #2}}}
\nc{\ddd}[2]{{{{\partial}^2 #1}\over{\partial {#2}^2}}}
\nc{\dddd}[3]{{{{\partial}^2 #1}\over
	{\partial #2 \partial #3}}}
\nc{\dder}[2]{{{d #1}\over{d #2}}}
\nc{\ddder}[2]{{{d^2 #1}\over{d {#2}^2}}}
\nc{\dddder}[3]{{d^2 #1}\over
	{d #2 d #3}}
\nc{\dx}[1]{d\,^{#1}x}
\nc{\dy}[1]{d\,^{#1}y}
\nc{\dz}[1]{d\,^{#1}z}
\nc{\dl}[1]{\frac{d\,^{#1}l}{(2\pi)^{#1}}}
\nc{\dk}[1]{\frac{d\,^{#1}k}{(2\pi)^{#1}}}
\nc{\dq}[1]{\frac{d\,^{#1}q}{(2\pi)^{#1}}}

\nc{\cc}{\mbox{$c.c.$ }}
\nc{\hc}{\mbox{$h.c.$ }}
\nc{\cf}{cf.\ }
\nc{\erfc}{{\rm erfc}}
\nc{\Tr}{{\rm Tr\,}}
\nc{\tr}{{\rm tr\,}}
\nc{\pol}{{\rm pol}}
\nc{\sign}{{\rm sign}}
\nc{\bfT}{{\bf T }}

\def\GeV{{\rm\ GeV}}

\nc{\cA}{{\cal A}}
\nc{\cB}{{\cal B}}
\nc{\cD}{{\cal D}}
\nc{\cE}{{\cal E}}
\nc{\cG}{{\cal G}}
\nc{\cH}{{\cal H}}
\nc{\cL}{{\cal L}}
\nc{\cO}{{\cal O}}
\nc{\cT}{{\cal T}}
\nc{\cN}{{\cal N}}
\nc{\rvac}[1]{|{\cal O}#1\rangle}
\nc{\lvac}[1]{\langle{\cal O}#1|}
\nc{\rvacb}[1]{|{\cal O}_\beta #1\rangle}
\nc{\lvacb}[1]{\langle{\cal O}_\beta #1 |}
\nc{\bb}{\bar{\beta}}
\nc{\bt}{\tilde{\beta}}
\nc{\ctH}{\tilde{\cal H}}
\nc{\chH}{\hat{\cal H}}
\nc{\1}{\aa}
\nc{\2}{\"{a}}
\nc{\3}{\"{o}}
\nc{\4}{\AA}
\nc{\5}{\"{A}}
\nc{\6}{\"{O}}
\nc{\al}{\alpha}
\nc{\g}{\gamma}
\nc{\Del}{\Delta}
\nc{\e}{\epsilon}
\nc{\eps}{\epsilon}
\nc{\lam}{\lambda}
\nc{\om}{\omega}
\nc{\Om}{\Omega}
\nc{\ve}{\varepsilon}
\nc{\mn}{{\mu\nu}}
\nc{\k}{\kappa}
\nc{\vp}{\varphi}

\nc{\advp}[3]{{\it  Adv.\ in\ Phys.\ }{{\bf #1} {(#2)} {#3}}}
\nc{\annp}[3]{{\it  Ann.\ Phys.\ (N.Y.)\ }{{\bf #1} {(#2)} {#3}}}
\nc{\apl}[3]{{\it  Appl. Phys. Lett. }{{\bf #1} {(#2)} {#3}}}
\nc{\apj}[3]{{\it  Ap.\ J.\ }{{\bf #1} {(#2)} {#3}}}
\nc{\apjl}[3]{{\it  Ap.\ J.\ Lett.\ }{{\bf #1} {(#2)} {#3}}}
\nc{\app}[3]{{\it Astropart.\ Phys.\ }{{\bf #1} {(#2)} {#3}}}
\nc{\cmp}[3]{{\it  Comm.\ Math.\ Phys.\ }{{ \bf #1} {(#2)} {#3}}}
\nc{\cqg}[3]{{\it  Class.\ Quant.\ Grav.\ }{{\bf #1} {(#2)} {#3}}}
\nc{\epl}[3]{{\it  Europhys.\ Lett.\ }{{\bf #1} {(#2)} {#3}}}
\nc{\ijmp}[3]{{\it Int.\ J.\ Mod.\ Phys.\ }{{\bf #1} {(#2)} {#3}}}
\nc{\ijtp}[3]{{\it Int.\ J.\ Theor.\ Phys.\ }{{\bf #1} {(#2)} {#3}}}
\nc{\jmp}[3]{{\it  J.\ Math.\ Phys.\ }{{ \bf #1} {(#2)} {#3}}}
\nc{\jpa}[3]{{\it  J.\ Phys.\ A\ }{{\bf #1} {(#2)} {#3}}}
\nc{\jpc}[3]{{\it  J.\ Phys.\ C\ }{{\bf #1} {(#2)} {#3}}}
\nc{\jap}[3]{{\it J.\ Appl.\ Phys.\ }{{\bf #1} {(#2)} {#3}}}
\nc{\jpsj}[3]{{\it J.\ Phys.\ Soc.\ Japan\ }{{\bf #1} {(#2)} {#3}}}
\nc{\lmp}[3]{{\it Lett.\ Math.\ Phys.\ }{{\bf #1} {(#2)} {#3}}}
\nc{\mpl}[3]{{\it  Mod.\ Phys.\ Lett.\ }{{\bf #1} {(#2)} {#3}}}
\nc{\ncim}[3]{{\it  Nuov.\ Cim.\ }{{\bf #1} {(#2)} {#3}}}
\nc{\np}[3]{{\it  Nucl.\ Phys.\ }{{\bf #1} {(#2)} {#3}}}
\nc{\npps}[3]{{\it  Nucl.\ Phys.\ Proc.\ Suppl.\ }{{\bf #1} {(#2)} {#3}}}
\nc{\pr}[3]{{\it Phys.\ Rev.\ }{{\bf #1} {(#2)} {#3}}}
\nc{\pra}[3]{{\it  Phys.\ Rev.\ A\ }{{\bf #1} {(#2)} {#3}}}
\nc{\prb}[3]{{\it  Phys.\ Rev.\ B\ }{{{\bf #1} {(#2)} {#3}}}}
\nc{\prc}[3]{{\it  Phys.\ Rev.\ C\ }{{\bf #1} {(#2)} {#3}}}
\nc{\prd}[3]{{\it  Phys.\ Rev.\ D\ }{{\bf #1} {(#2)} {#3}}}
\nc{\prl}[3]{{\it Phys.\ Rev.\ Lett.\ }{{\bf #1} {(#2)} {#3}}}
\nc{\pl}[3]{{\it  Phys.\ Lett.\ }{{\bf #1} {(#2)} {#3}}}
\nc{\prep}[3]{{\it Phys.\ Rep.\ }{{\bf #1} {(#2)} {#3}}}
\nc{\prsl}[3]{{\it Proc.\ R.\ Soc.\ London\ }{{\bf #1} {(#2)} {#3}}}
\nc{\ptp}[3]{{\it  Prog.\ Theor.\ Phys.\ }{{\bf #1} {(#2)} {#3}}}
\nc{\ptps}[3]{{\it  Prog\ Theor.\ Phys.\ suppl.\ }{{\bf #1} {(#2)} {#3}}}
\nc{\physa}[3]{{\it  Physica\ A\ }{{\bf #1} {(#2)} {#3}}}
\nc{\physb}[3]{{\it  Physica\ B\ }{{\bf #1} {(#2)} {#3}}}
\nc{\phys}[3]{{\it Physica\ }{{\bf #1} {(#2)} {#3}}}
\nc{\rmp}[3]{{\it  Rev.\ Mod.\ Phys.\ }{{\bf #1} {(#2)} {#3}}}
\nc{\rpp}[3]{{\it Rep.\ Prog.\ Phys.\ }{{\bf #1} {(#2)} {#3}}}
\nc{\sjnp}[3]{{\it Sov.\ J.\ Nucl.\ Phys.\ }{{\bf #1} {(#2)} {#3}}}
\nc{\spjetp}[3]{{\it Sov.\ Phys.\ JETP\ }{{\bf #1} {(#2)} {#3}}}
\nc{\yf}[3]{{\it Yad.\ Fiz.\ }{{\bf #1} {(#2)} {#3}}}
\nc{\zetp}[3]{{\it Zh.\ Eksp.\ Teor.\ Fiz.\  }{{\bf #1}  {(#2)} {#3}}}
\nc{\zp}[3]{{\it Z.\ Phys.\ }{{\bf #1} {(#2)} {#3}}}
\nc{\ibid}[3]{{\sl ibid.\ }{{\bf #1} {#2} {#3}}}
\nc{\rf}[1]{(\ref{#1})}
\nc{\nn}{\nonumber \\*}
\nc{\bfB}{\bf{B}}
\nc{\bfv}{\bf{v}}
\nc{\bfx}{\bf{x}}
\nc{\bfy}{\bf{y}}
\nc{\vx}{\vec{x}}
\nc{\vy}{\vec{y}}
\nc{\oB}{\overline{B}}
\nc{\oI}{\overline{I}}
\nc{\oR}{\overline{R}}
\nc{\rar}{\rightarrow}
\nc{\ti}{\times}
\nc{\slsh}{\hskip-5pt/}
\nc{\sm}{Standard~Model~}
\nc{\MP}{M_{\rm Pl}}
\nc{\tp}{t_{\rm Pl}}
\nc{\ave}{\bar{E}}

\nc{\eff}{{\rm eff}}
\nc{\kk}{\vek{k}}
\nc{\pp}{{\rm p}}
\nc{\ga}{g_{a\gamma}}
\nc{\vv}{\\}
\nc{\eee}{{\bf E}}
\nc{\bbb}{{\bf B}}
\nc{\qcd}{T_{\rm QCD}}
\nc{\G}{\rm \ G}
\def\vec#1{{\bf #1}}

\def\lae{\;^{<}_{\sim} \;} \def\gae{\; ^{>}_{\sim} \;} 

\begin{document}
{\title{\vskip-2truecm{\hfill {{\small \\
	\hfill GUTPA/99/XX/X\\
	}}\vskip 1truecm}
{\bf Positive Order $H^2$ Mass Corrections and Affleck-Dine Baryogenesis}}
{\author{
{\sc  John McDonald$^{1}$}\\
{\sl\small Department of Physics and Astronomy, University of Glasgow,
Glasgow G12 8QQ, SCOTLAND}
}
\maketitle
\begin{abstract}
\noindent

      It is usually assumed that the order $H^2$ corrections to 
the SUSY-breaking mass squared terms in the early Universe must
be negative in order to allow the Affleck-Dine mechanism to work.
We reconsider this assumption in the context of D-term inflation models
for the case where the mass squared
term has a correction $cH^2$ with $c>0$. We show that, in general, the 
baryon asymmetry is likely to be too small if $c > 9/16$. However,
for $c$ as large as 0.5
the observed baryon asymmetry can be readily generated; in
particular, for $d=6$ directions
the observed asymmetry can be produced for a wide range of
reheating temperatures, in contrast with the case of 
negative $H^2$ corrections which require a reheating
temperature around $1 \GeV$. 
Thus positive $H^2$ corrections do not rule out the Affleck-Dine
mechanism and can even greatly broaden its applicability to inflation models.

\end{abstract}
\vfil
\footnoterule
{\small $^1$mcdonald@physics.gla.ac.uk}

\thispagestyle{empty}
\newpage
\setcounter{page}{1}


\section{Introduction}

      Affleck-Dine (AD) baryogenesis \cite{ad} is a
natural possibility for generating
the baryon asymmetry in the context of the MSSM \begin{footnote}{
For a recent review of baryogenesis, see \cite{riotrod}.}\end{footnote}. It is based on inducing
a baryon asymmetry in a coherently oscillating squark condensate. In the
modern view of the AD mechanism \cite{drt,drt2},
the initial value of the AD field and its
subsequent evolution depends crucially on the order $H^2$ corrections
to the SUSY breaking mass squared terms that are expected once the
energy density in the early Universe and non-minimal K\"ahler couplings are taken into account \cite{drt,h2}. The natural
assumption to make is that such corrections must have a negative sign, 
so that the minimum of the AD potential is non-zero at early times
when the expansion rate is large, turning positive (and so allowing the AD
scalar to coherently oscillate) only once the gravity-mediated SUSY
breaking terms come to dominate. Positive order $H^2$ mass corrections are commonly 
thought to result in an AD field with an amplitude too small to account
for the B asymmetry. However, it is not obvious that this is generally true
and it is this issue that we wish to explore here.
The case for negative order $H^2$ corrections was originally put forward in the context of F-term inflation models, under the assumption that
the sign of the $H^2$ corrections during and
after inflation are the same \cite{drt2}. In this case, if
the $H^2$ terms were positive, 
the AD scalar would be exponentially damped nearly to zero
during inflation. It was later assumed \cite{kmr} that
the same condition had to hold for D-term inflation
models \cite{dti}. However, these have no order $H^2$ corrections
to scalar masses along D-flat directions during inflation.
We will show that in D-term inflation models with
positive order $H^2$ corrections (and also in F-term
inflation models which have a change in sign of the
$H^2$ correction from negative to positive after inflation) it is
possible to have successful AD baryogenesis, with a much wider applicability to 
inflation models than in the case of negative $H^2$ corrections.

       In the AD mechanism, a baryon asymmetry is
induced when the A-term in the scalar potential produces a phase
shift between the real and imaginary parts of the AD field \cite{sewm}.
In D-term inflation models, the A-term receives
no order $H$ correction either during or after inflation \cite{kmr,dti}.
Thus the
initial phase of the AD scalar is random 
and so is typically of the order of one relative to the real direction.
The final asymmetry will be
fixed once $ H \lae m$, where $m \approx 100\GeV$
is the gravity-mediated SUSY breaking scalar mass term, at which point the
B violating terms become suppressed relative to the B conserving terms
in the potential. Thus in order to calculate the baryon asymmetry we need
to calculate the amplitude 
and the phase shift of the AD field at $H \approx m$.

\section{Amplitude at H $\approx$ m.}

                 The scalar potential of the AD field along
a dimension $d$ flat direction has the general form \cite{drt2},
\be{e1} V = (m^2 + c H^2) |\Phi|^2
+ \left( \frac{A_{\lambda} \lambda \Phi^d}{d M^{d-3}}
+ h.c.\right) + \frac{\lambda^2 |\Phi|^{2(d-1)}}{M^{2(d-3)}}    ~,\ee
where $M = M_{Pl}/\sqrt{8 \pi}$ and typically $\lambda \sim 1/(d-1)!$ for Planck-scale
non-renormalizable terms. We will focus on R-parity conserving
models, for which $d$ is even \cite{drt2}, and consider the cases $d=4$ and $d=6$.
The evolution of the amplitude
depends on whether the $|\Phi|^2$ term or the non-renormalizable term
is dominant, with the value of the AD scalar below which the
potential is $\phi^2$ dominated, $\phi_{c}$,
 being given by
\be{e2} \phi_{c} = \sqrt{2}
\left(\frac{c}{\lambda^{2}}\right)^{\frac{1}{2(d-2)}}
\left(H^2 M^{2(d-3)}\right)^{\frac{1}{2(d-2)}}       ~,\ee
where $|\Phi| = \phi/\sqrt{2}$. 
\newline \underline{Evolution for $\phi < \phi_{c}$:} When
$\phi < \phi_{c}$
the scalar field evolves according to
\be{e3} \ddot{\phi} + 3 H \dot{\phi} = - c H^2 \phi         ~.\ee
After inflation (but prior to reheating),
the Universe is matter dominated by inflaton oscillations.
For a matter dominated Universe \eq{e3} has the solution
\be{e4}   \phi  \propto a^{-\eta} \;\;\;\;;\;\;
\eta = \frac{1}{2} \left[ \frac{3}{2}
- \sqrt{ \frac{9}{4} - 4 c} \right]         ~,\ee
where $a$ is the scale factor.
There are then two distinct cases. If $c > 9/16 \approx 0.56$,
the square root is imaginary and we have a
damped oscillating solution of the form
\be{e4a} \phi =  \left(\frac{a_{o}}{a}\right)^{3/4} \phi_{o}
{\rm Cos} \left( \gamma {\rm log} \left(\frac{a_{o}}{a}\right) \right) \;\;\; ; \;\;\; \gamma 
= \frac{1}{2} \sqrt{4c-\frac{9}{4}}        ~.\ee
Note that the amplitude falls off less rapidly than the corresponding amplitude for 
a coherently oscillating scalar with a constant mass term, for which $\phi \propto a^{-3/2}$. 
If, on the other hand, $c < 9/16$, then the field
is damped but not oscillating, with a magnitude given
by 
\be{e4b} \phi = \left(\frac{a_{o}}{a}\right)^{\eta}\phi_{o}     ~.\ee
\underline{Evolution for $\phi > \phi_{c}$:} For $\phi > \phi_{c}$, the field is effectively oscillating in
 a $\phi^{2(d-1)}$ potential. The energy density evolves as \cite{turner}
$\rho \propto a^{-6(d-1)/d}$
and so the amplitude of the oscillation evolves as
 \be{e5} \phi \propto a ^{- \frac{3}{d}}             ~.\ee
Thus for a field which starts oscillating at $\phi_{o} > \phi_{c}$,
the amplitude at $m \lae H < H_{c}$ is given by (from now we 
use $\eta \equiv Re(\eta)$)
\be{e6} \phi = \left(\frac{H}{H_{c}}\right)^{\frac{2 \eta}{3}} 
\left(\frac{H_{c}}{H_{o}}\right)^{\frac{2}{d}} \phi_{o}    ~,\ee
where $H_{c}$ is the value at which $\phi = \phi_{c}$,
\be{e6a} H_{c}^{4-d} = 
\left(\frac{\lambda^2}{c}\right)^{d/2} 
\left(\frac{1}{H_{o}}\right)^{2(d-2)} 
\left(\frac{1}{M}\right)^{d(d-3)} 
\left(\frac{\phi_{o}}{\sqrt{2}}\right)^{d(d-2)} 
~\ee
and $H_{o}$ is the value at the end of inflation. For the case $\phi_{o} < \phi_{c}$, 
$H_{c} \rightarrow H_{o}$ in \eq{e6}.

     For the $d=4$ case $H_{c}$ is not fixed by \eq{e6a}, because
both $\phi$ and $\phi_{c}$ for a given value of  $H$ are proportional to
$H^{1/2}$.
Thus if the value of $\phi$ at the end of inflation,
$\phi_{o}$, is less than $\phi_{c}$ then it will remain
so until $H \approx m$. $\phi_{c}$ at $H_{o}$ is given by
\be{e7} \phi_{c} =  7 \times 10^{15} 
\left(\frac{c}{\lambda^2}\right)^{1/4} 
\left(\frac{H_{o}}{10^{13} \GeV}\right)^{1/2} \GeV      ~,\ee
where we are considering the typical value at the end of
inflation to be $H_{o} \approx
10^{13} \GeV$, in accordance with the observed 
cosmic microweave background perturbations \cite{cmb,lr}.
For the $d=6$ case, $H_{c}$ and $\phi_{c}$ are given by
\be{e9} H_{c} = 1.7 \times 10^{27} 
\left(\frac{c}{\lambda^2}\right)^{3/2}
\left(\frac{H_{o}}{10^{13} \GeV}\right)^{4}
\left(\frac{10^{16} \GeV}{\phi_{o}}\right)^{12}
\GeV
~\ee
and
\be{e10} \phi_{c} =  5.5 \times 10^{17} 
\left(\frac{c}{\lambda^2}\right)^{1/2}
\left(\frac{H_{o}}{10^{13} \GeV}\right)
\left(\frac{10^{17} \GeV}{\phi_{o}}\right)^{3}
\GeV
~.\ee 

       For example, consider the case with
the largest degree of damping, $c > 9/16 \approx 0.56$.
In this case the amplitude of the oscillations
evolves as $\phi \propto a^{-3/4}$.
We will see that $\phi_{o} \lae \phi_{c}$ must
be satisfied in order to have acceptable
energy density perturbations in both the $d=4$ and $d=6$ cases. 
In this case the field at $H \approx m$ is generally given by 
\be{e8} \phi \left(H \approx m\right) \approx 3 \times 10^{10} 
\left(\frac{10^{13} \GeV}{H_{o}}\right)^{1/2}
\left(\frac{\phi_{o}}{10^{16} \GeV}\right) \GeV    ~.\ee
Thus the amplitude of the AD field at $H \approx m$ is
not very small so long as the initial value of the AD field
at the end of inflation is large enough. In fact, the amplitude of \eq{e8} 
is typically of the order of the amplitude expected in the $d=4$ case with 
negative $H^2$ corrections, $\phi \approx (Mm)^{1/2} \approx 10^{10} \GeV$.
Therefore so long as the phase shift is sufficiently large 
we should be able to generate the observed asymmetry.

\section{Phase Shift and Baryon Asymmetry}

      In general, the phase shift will be the product of an initial CP
 violating phase and a "kinematical factor" coming
 from the time evolution of the real and imaginary components of the
 AD field. Assuming an initial phase of order one, the phase shift
 $\delta$ will be given by 
\be{e12}  \delta  \approx  \int \Delta m \; dt  ~,\ee
where $\Delta m$ is the difference in the effective mass terms
for the scalars along the real and imaginary
directions of the AD field ($\Phi \equiv (\phi_{R} + i \phi_{I})/\sqrt{2}$).
We will consider the
phase shift for $\phi_{o} \lae \phi_{c}$; it is straightforward to
show that the contribution to the phase
shift from $\phi > \phi_{c}$ is generally negligible compared with this. 

From the equations of motion for
$\phi_{R}$ and $\phi_{I}$, there is a shift in the effective
mass squared term,
$m_{eff}^2 \rightarrow m_{eff}^2 \pm \delta m^2/2$ due to the
A-term contribution, where 
$m_{eff}^2 \equiv \partial^{2} V(\phi) /2 \partial \phi^{2}$ and where
\be{e13} \delta m^2 \approx \frac{A_{\lambda} \lambda}{2^{d/2-1}}
\frac{\phi_{o}^{d-2}}{M^{d-3}} \left(\frac{H}{H_{o}}\right)^{\frac{2 \eta}{3}(d-2)}
~.\ee
For small $\delta m^2$ we therefore have 
\be{e13a}  \Delta m \approx \delta m^2/2m_{eff}
= K_{d} H^{\frac{2\eta}{3}(d-2)-1}   ~,\ee
where  
\be{e13b} K_{d} = 
\frac{A_{\lambda} \lambda}{2^{d/2} c^{1/2}}
\frac{\phi_{o}^{d-2}}{M^{d-3}}
\left(\frac{1}{H_{o}}\right)^{\frac{2 \eta}{3}(d-2)}
~.\ee
Thus the phase shift is given by
\be{e14} \delta = \frac{2 K_{d}}{3 \left(\frac{2 \eta}{3}(d-2)-2)\right)} 
(H_{o}^{\frac{2 \eta}{3}(d-2)-2}
- H^{\frac{2 \eta}{3}(d-2)-2})      ~.\ee
In general this will be dominated by the contribution at $H \approx m$ 
\begin{footnote}{\eq{e14} is strictly correct for the case $c > 9/16$, with 
$\phi$ oscillating throughout. For $c < 9/16$ the field only begins oscillating 
once $H \approx m$. However, since most of the phase shift arises at this epoch, 
\eq{e14} is essentially correct for this case also.}\end{footnote}.
The B asymmetry density at $H \approx m$ is then
$n_{B} \approx m \phi^2 Sin \delta$.
This is related to the present baryon-to-photon ratio $\eta_{B}$ by 
\be{e14a} n_{B} = \left(\frac{\eta_{B}}{2 \pi}\right)  \frac{H^2 M_{Pl}^2}{T_{R}}   
~.\ee
We next calculate $\eta_{B}$ for the $d=4$ and $d=6$ cases.
\newline \underline{$d=4$:} In this case the phase shift is given by 
\be{e151} \delta \approx 0.7 
\left(\frac{\lambda}{2-4 \eta/3}\right)
\left(\frac{10^{13} \GeV}{H_{o}}\right)^{4 \eta/3} 
\left(\frac{\phi_{o}}{10^{16} \GeV}\right)^{2} 
\left(10^{11}\right)^{1 - 4\eta/3} 
~.\ee
where we have taken $A_{\lambda} \approx m \approx 100 \GeV$. 
Thus typically the phase is of order one. 
$\eta_{B}$ is then given by 
\be{e152} \left( \frac{\eta_{B}}{10^{-10}}\right) \approx 40 
\left(\frac{10^{13} \GeV}{H_{o}}\right)^{4 \eta/3} 
\left(\frac{\phi_{o}}{10^{16} \GeV}\right)^{2} 
\left(\frac{T_{R}}{10^9 \GeV}\right)
\left(10^{11}\right)^{1 - 4 \eta/3} 
~.\ee
For the case $c > 9/16$ (corresponding to $\eta = 3/4$) 
and with $\phi_{o} \approx 10^{16} \GeV$ (the upper limit from 
scale-invariant density perturbations, as discussed below) this is marginally 
able to account for the observed B asymmetry 
($\eta_{B} \approx (3-8) \times 10^{-10}$ \cite{sarkar})
if $T_{R} \gae 10^{8} \GeV$, close to the upper bound from 
thermal gravitino production, $T_{R} \lae 10^{9} \GeV$.
As discussed in detail for the $d=6$ case below, the asymmetry
 can be enhanced (corresponding to 
lower values of $T_{R}$ being compatible with the observed B asymmetry) if $c$ is 
even slightly less than $9/16$; for example, $c = 0.5$ gives a $5 \times 10^3$ enhancement of the
B asymmetry. However, in the context of D-term inflation models, we will
see that the $d=4$ condensate,
as in the case of negative order $H^2$ corrections \cite{kmr},
is unlikely to survive thermalization by the first stage of
reheating after inflation. 
\newline \underline{$d=6$:}  In this case 
the phase shift is given by 
\be{e153} \delta \approx 10^{-16}  \left( \frac{\lambda}{2-8 \eta/3}\right)
\left(\frac{10^{13} \GeV}{H_{o}}\right)^{8 \eta/3} 
\left(\frac{\phi_{o}}{10^{16} \GeV}\right)^{4} 
\left(10^{11}\right)^{2-8 \eta/3} 
~.\ee
Thus typically the phase shift is no larger than
around $10^{-10}$ for $c > 9/16$. (This is much smaller than 
in the case of negative $H^2$ corrections, for which $\delta \approx 1$.)
$\eta_{B}$ is then given by
\be{e15} \left(\frac{\eta_{B}}{10^{-10}}\right)
\approx 3 \times 10^{-16} \left(\frac{\lambda}{2-8 \eta/3}\right)
\left(\frac{10^{13} \GeV}{H_{o}}\right)^{4 \eta} 
\left(\frac{\phi_{o}}{10^{16} \GeV}\right)^{6} 
\left(\frac{T_{R}}{10^9 \GeV}\right)
\left(10^{11}\right)^{3 - 4 \eta} 
~.\ee
From this we see that even with $\phi_{o} \approx 10^{17} \GeV$,
which we will show is the upper limit from the scale-invariance of the
density perturbation spectrum, the highly suppressed phase shift in 
the case $c > 9/16$ means that we 
 cannot produce a value of $\eta_{B}$
much above $10^{-20}$. Thus there is no possibility of accounting for the
observed B asymmetry
in this case. However, once $c < 9/16$ there is a rapid enhancement of the
B asymmetry.
This can be seen from the "enhancement
factor" $\left(10^{11}\right)^{3-4 \eta}$ in \eq{e15}. This gives the observed
asymmetry for $T_{R} \approx 10^{8} \GeV$ and $\phi_{o} \approx 10^{17} \GeV$ if 
$c = 0.5$. The corresponding values for other $\phi_{o}$ and $T_R$
are given in Table 1.
From this we see that, with values of
$c$ between 0.2 and 0.5, the observed asymmetry can be
accomodated by a wide range of reheating temperatures.
\newpage

\begin{center}{\bf Table 1. d=6 B asymmetry enhancement factor.
}\end{center}

\begin{center}
\begin{tabular}{|c|c|c|c|}          \hline
c & Enhancement factor & $T_{R}(\phi_{o} = 10^{17}\GeV)$ & 
$T_{R}(\phi_{o} = 10^{16}\GeV)$  \\ \hline
0.51 & $ 3 \times 10^{10}$ &  $10^{9} \GeV$ & $10^{15} \GeV$  
 \\
0.50 & $ 1 \times 10^{11}$ &  $10^{8} \GeV$ & $10^{14} \GeV$  
 \\
0.45 & $ 5 \times 10^{14}$ &  $10^{4} \GeV$ & $10^{10} \GeV$  
 \\
0.40 & $ 5 \times 10^{17}$ &  $10 \GeV$ & $10^{7} \GeV$  
 \\
0.35 & $ 2 \times 10^{20}$ &  $0.1 \GeV$ & $10^{5} \GeV$  
 \\
0.30 & $ 4 \times 10^{22}$ &  $10^{3} \GeV$ & $10^{3} \GeV$  
 \\
0.20 & $ 3 \times 10^{26}$ &  $10^{-7} \GeV$ & $0.1 \GeV$  
 \\
\hline
\end{tabular}
\end{center} 

   Therefore, although values of $c$ close to or larger than 1
   are indeed ruled out, for values as large as 0.5 the observed asymmetry can be  
generated along the d=6 direction. Such a value 
of $c$ seems entirely plausible and consistent with the expectation
that $c$ is of the order of one.
Indeed, the observed asymmetry can be accomodated by a much wider range of reheating
temperatures than in the case of negative $H^2$ corrections, which requires
 $T_{R}$ to be of the order of $1 \GeV$.

\section{Density Perturbation Constraints}
                     
                The magnitude of the B asymmetry depends on the initial value of the 
AD field. This cannot be too large during inflation, otherwise the energy density of 
the AD field will dominate the spectral index of the energy density perturbations, 
resulting in an unacceptable deviation from scale-invariance \cite{isobb}. This has already been 
discusssed for the $d=6$ case in the context of D-term inflation, where it was shown that the requirement that the spectral index is dominated by the inflaton field implies that \cite{isobb}
\be{e16} \phi \lae \phi_{max} \approx 0.48 \left(\frac{g}{\tilde{\lambda}}\right)^{1/4} 
\left (M \xi \right)^{1/2}  \lae 10^{17} \GeV   ~,\ee
where $g$ and $\tilde{\lambda}$ are the Fayet-Illiopoulos gauge coupling and superpotential 
coupling of the D-term inflation model and $\xi \approx 6.6 \times 10^{15} \GeV$ \cite{dti,lr}.
For the $d=4$ case the analogous requirement is that 
\be{e17} \phi \lae \phi_{max} \approx 0.23
\left(\frac{g}{\tilde{\lambda}}\right)^{1/2}\xi  < 10^{16} \GeV   ~,\ee
where in general $\tilde{\lambda}$ cannot be very small compared with $g$,
otherwise there
would be efficient parametric resonant decay of the inflaton and
violation of the thermal gravitino production
bound \cite{bbdti}. However, there is a generic problem for the 
$d=4$ direction in the context of D-term inflation models \cite{kmr}.
 D-term inflation models typically have a 
a first stage of reheating immediately after the end of inflation 
from the decay of the $\psi_{-}$ field of the inflaton sector, 
with $T_{R}^{\psi_{-}} \approx g^{1/2} \xi \approx 10^{15} \GeV$. In order to
avoid thermalization by this radiation density, the AD field at the end of inflation must
be greater than $ H_{o}^{3/8}M^{5/8} \approx 2 \times 10^{16} \GeV$ \cite{kmr}. Thus 
the upper bound from scale-invariance
implies that the $d=4$ directions with positive
$H^2$ corrections are thermalized,
as in the case of negative $H^2$ corrections.
The $d=6$ direction can, however, avoid thermalization and 
still be consistent with scale-invariance.

\section{F-term Inflation}

      So far we have been considering D-term inflation. However, the above
discussion will also apply to F-term inflation models
if some conditions are met. 
The main differences between D-term and F-term inflation are that the
AD scalar will receive order $H^2$ corrections to its mass both during and
after inflation and that the A-term will also receive order $H$
corrections. The effect of the order $H$ corrections to the
A-term is that the
initial CP violating phase will be set by the explicit CP violating phase
appearing in the A-term (which can be much smaller than one)
rather than by a random initial condition.
The effect of the positive order $H^2$ mass correction during inflation
is to exponentially drive the AD scalar to zero, so ruling out
AD baryogenesis \cite{drt2}.
However, it is still possible that the order $H^2$ term could become
positive {\it after} inflation, if there was a sign change of the $H^2$
correction at the end of inflation. This is possible, for example,
if the energy density after inflation was made up of two fields
with comparable energies. In this case the $H^2$ term would be the sum of
contributions from each field, and the sum of the contributions could change
as the energy is transferred away from the vacuum energy
at the end of inflation. Thus it is conceivable that the
$H^2$ correction could be negative
during inflation but positive after. The 
D-term discussion would then apply to this case also,
since the value of the fields at the end
of inflation from minimizing the potential during inflation
simply corresponds to $\phi_{c}$. 
\section{Conclusions}

    In D-term inflation models, the Affleck-Dine mechanism with a positive $cH^2$
correction to the SUSY breaking mass squared term can
generate the observed B asymmetry if
$c \lae 0.5$. The $d=4$ direction is ruled out by the near 
scale-invariance of the density perturbations together with the
thermalization of the AD field during the first stage of
reheating. The $d=6$ direction, on the other hand, can
successfully generate the observed
baryon asymmetry, and for a much wider range of
reheating temperatures than in the
case of negative $H^2$ corrections. Similar conclusions apply to the case
of F-term inflation models with a positive $H^2$ correction after inflation.
Therefore the AD mechanism can have
a much wider range of applicability to particle physics
models than previously thought.
In particular, since the reheating temperature in $d=4$ and $d=6$
models with a CP violating phase $\approx 1$ is constrained to be very high
or very low ($10^8 \GeV$ and $1 \GeV$ respectively), in many inflation models
the AD mechanism with positive $H^2$ corrections will be the natural option.

\subsection*{Acknowledgements}   This work has been supported by the PPARC.

\end{document}